\documentclass[a4paper, conference]{IEEEtran}
\IEEEoverridecommandlockouts
\usepackage{cite}
\usepackage{amsmath,amssymb,amsfonts}
\usepackage{algorithmic}
\usepackage{graphicx}
\usepackage{textcomp}
\usepackage{xcolor}
\def\BibTeX{{\rm B\kern-.05em{\sc i\kern-.025em b}\kern-.08em
    T\kern-.1667em\lower.7ex\hbox{E}\kern-.125emX}}
\usepackage[left=1.3cm,right=1.3cm,top=1.9cm,bottom=4.3cm]{geometry}
\makeatletter
\def\hlinewd#1{%
\noalign{\ifnum0=`}\fi\hrule \@height #1 %
\futurelet\reserved@a\@xhline}
\makeatother

\begin{document}
\title{Virtio-FPGA: a virtualization solution for SoC-attached FPGAs \\
\thanks{ This work has received funding from the EU Horizon 2020 Programme under grant agreement No 957269 (EVEREST) \cite{b15}. }
}
\author{\IEEEauthorblockN{Anna Panagopoulou}
\IEEEauthorblockA{\textit{Virtual Open Systems SAS} \\
Grenoble, France \\
anna@virtualopensystems.com}
\and
\IEEEauthorblockN{Michele Paolino}
\IEEEauthorblockA{\textit{Virtual Open Systems SAS} \\
Grenoble, France \\
m.paolino@virtualopensystems.com}
\and
\IEEEauthorblockN{Daniel Raho}
\IEEEauthorblockA{\textit{Virtual Open Systems SAS} \\
Grenoble, France \\
s.raho@virtualopensystems.com}
}

\maketitle

\begin{abstract}
Recently, FPGA accelerators have risen in popularity as they present a suitable way of satisfying the high-computation and low-power demands of real time applications. The modern electric transportation systems (such as aircraft, road vehicles) can greatly profit from embedded FPGAs, which incorporate both high-performance and flexibility features into a single SoC. At the same time, the virtualization of FPGA resources aims to reinforce these systems with strong isolation, consolidation and security.
In this paper, we present a novel virtualization framework aimed for SoC-attached FPGA devices, in a Linux and QEMU/KVM setup.
We use Virtio as a means to enable the configuration of FPGA resources from guest systems in an efficient way. Also, we employ the Linux VFIO and Device Tree Overlays technologies in order to render the FPGA resources dynamically accessible to guest systems.
The ability to dynamically configure and utilize the FPGA resources from a virtualization environment is described in details. The evaluation procedure of the solution is presented and the virtualization overhead is benchmarked as minimal (around 10\%) when accessing the FPGA devices from guest systems.
\end{abstract}

\begin{IEEEkeywords}
FPGAs, virtio, FPGA Manager, VFIO
\end{IEEEkeywords}

\section{Introduction}

In the recent years, SoC-attached FPGAs has emerged as an architecture of broad interest in the electric transportation systems\cite{b1}.
It is a formation that comes with both performance and flexibility gains, as it combines general-purpose and highly-specialized computation in a single board \cite{b2}. 
At the same time, virtualizing the FPGA resources of SoC-attached FPGAs is the means to build standardized interfaces to access the reconfigurable, heterogeneous hardware.

The virtualization of FPGA resources is a non-trivial goal, due to the unique nature of FPGAs; the hardware is modifiable and introduces great challenges in exposing a unified layer for the virtual resources \cite{b3}. The objective of the proposed solution is to address some of the FPGA virtualization challenges of ARM SoCs, on the grounds of well-known and accepted APIs and environments. In this paper, we are going to present Virtio-FPGA; an initial prototype built for Linux and QEMU/KVM, that aims to provide transparent and dynamic accesses of the FPGA hardware to guest systems.

The rest of this paper is organized as follows: Section II
covers the background knowledge about the involved technologies. Then, Section III describes the technical details of the Virtio-FPGA solution, while Section IV presents the evaluation environment and the obtained results. Section V outlines related works and Section VI concludes the paper, with discussion on the future directions.

\section{Background}

In this section we will present the basic concepts and ideas of the Linux technologies involved in the Virtio-FPGA solution. 
We will first provide background knowledge on the Virtio standard, then we will present the Linux FPGA Manager component of the Linux kernel and finally we will provide background knowledge on the VFIO pass-through for ARM in QEMU and the Device Tree Overlays technologies.

\subsection{Virtio}
Virtio \cite{b4} is an I/O virtualization standard, that defines an abstraction layer to enable communication between guests and host in a paravirtualized fashion. 

The Virtio standard is commonly implemented in the QEMU hypervisor, for emulating paravirtualized devices. QEMU exposes a virtual device to the guest, which can be driven by a front-end driver residing in the guest system. At the other end, the back-end driver is part of QEMU and is responsible for emulating the device, by interfacing with the host system.

In the base of the Virtio communication, there exists the virtqueue abstraction. In simple terms, the virtqueues represent memory regions of the guest system, that are shared with the host. For what concerns the exchange of data, the guest is responsible to allocate the memory regions as buffers that will be consumed by the host, and then configure the virtqueues to point to those buffers. So, the buffers become shared between the two parties and communication is made possible.

Performance-wise, Virtio is a high-efficient solution to accomplish guest-host communication, precisely because it is depending on a shared memory model.

\subsection{The Linux FPGA Manager}
The Linux FPGA Manager API \cite{b5} is a set of kernel functions with the aim of programming during the runtime  an FPGA device with a hardware description image. The API is implemented as a driver that conceals all the manufacturer-specific FPGA programming details from the upper software layers. 
In particular, the idea is that the manufacturer-related procedures are part of underlying drivers, that register themselves to the generic FPGA Manager API. Each of these drivers has to implement a set of operations, so that the interface with the generic API is properly setup. To mention the most basic of them, a \texttt{write\_init} operation has the role of preparing the FPGA to receive data, a \texttt{write} operation writes a buffer with data to the FPGA and a \texttt{write\_complete} is triggered after all the image data have been written.

\begin{figure}[htbp]
\centerline{\includegraphics[scale=0.4]{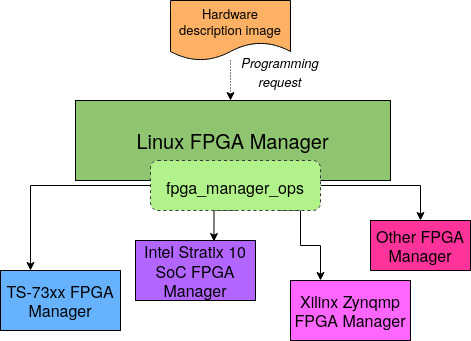}}
\caption{The Linux FPGA Manager as an interface to hide manufacturer-specific details of the FPGA programming operation.}
\label{fig:fig1}
\end{figure}

The FPGA Manager supports programming FPGAs using binary data in the form of either a scatter gather list, a single contiguous buffer, or a firmware file. All the parsing and interpretation of these vendor exclusive binary data, is to be performed by the underlying drivers. The relationship between the generic Linux FPGA Manager core and the underlying drivers is illustrated in figure \ref{fig:fig1}.

\subsection{QEMU VFIO pass-through for ARM}
VFIO pass-through \cite{b6} is a method that allows devices sitting behind an IOMMU controller to become directly available to guest systems. In a Linux environment, QEMU is capable of supporting VFIO pass-through for both PCI and platform (e.g. part of an SoC) devices.

VFIO is a Linux API \cite{b7} which permits virtual memory addresses to be used for managing physical devices. In its base, VFIO interfaces with the IOMMU subsystem in order to setup specific memory translations, from virtual to physical memory regions. Consequently, user-space components gain the ability to program and utilize the exposed device, after proper configuration is done.
Thus, QEMU, being a user-space software layer, becomes capable of driving the exposed device to make it accessible to guest systems.

To further detail the role of QEMU, it uses the VFIO API in a way that for a given device, specific host virtual addresses are mapped to device-accessible physical addresses. These host virtual addresses are basically guest physical addresses; therefore, the guest drivers can seemingly program the device given guest physical addresses, although these are transparently mapped to the actual physical memory regions.

\begin{figure}[htbp]
\centerline{\includegraphics[scale=0.17]{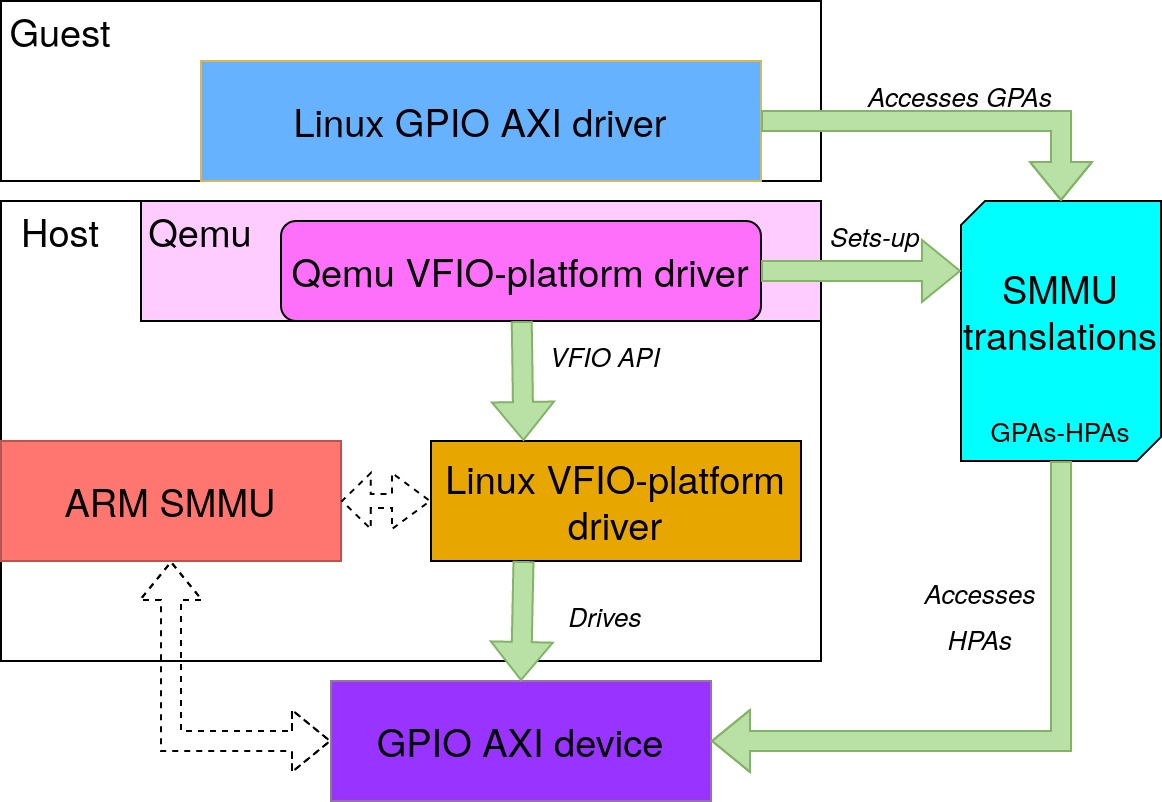}}
\caption{QEMU VFIO-passthrough technology for an example GPIO AXI device.}
\label{fig:fig4}
\end{figure}

For embedded ARM FPGAs, the solution to achieve direct pass-through of FPGA devices to guest systems is by utilizing QEMU's VFIO platform driver. This QEMU solution can target devices that are bound to Linux VFIO platform driver \cite{b8}, and thus, are available for user-space driving. In order for this exposure to be done safely, the candidate device should be part of the IOMMU subsystem, namely the SMMU on ARM SoCs. Figure \ref{fig:fig4} demonstrates the components of the VFIO pass-through technology, as provided by the QEMU hypervisor, for an example GPIO device present on the AXI subsystem.

\subsection{Device Tree overlays}

The device tree overlays are blobs that can modify the current, live device tree of the Linux kernel. The dynamic application of overlays can be enabled in a Linux kernel through the \texttt{CONFIG\_OF} compilation option. Also, a patch to enable \texttt{CONFIGFS} support for applying the overlays from a user-space context is required in the kernel. 

In general, the re-configuration of FPGAs during runtime, can greatly benefit from device tree overlays support. Given a hardware configuration image and also a corresponding device tree blob description, not only the FPGAs can be re-configured, but also the configured devices can be made visible to the system on-the-go.

\section{Virtio-FPGA}

In this section, we introduce the technical aspects of the Virto-FPGA solution for SoC-attached FPGAs.
Virto-FPGA is an ARM virtualization solution based on the QEMU/KVM hypervisor, that unveils the configuration and utilization of FPGA resources to Linux virtual machines.
It is an ensemble of software components which provide a dynamic way for virtual machines to program, then recognize and interface with the devices that constitute the re-configurable FPGA hardware. 

In the Virtio-FPGA solution, the virtualization challenge is addressed in a two-fold way: the FPGAs configuration and the FPGA resources utilization from virtual machines. With this design approach, we aim at reducing the complexity of the problem by exploiting existing Linux solutions to the greatest extent.

Specifically, the Virtio-FPGA solution is based on the Virtio standard for the virtual machines to interface with the Linux FPGA Manager component of the host system. This way, virtual machines become able to dynamically dispatch configuration requests to FPGA devices. Then, using proper orchestration of VFIO and Device Tree Overlays technologies, the Virtio-FPGA solution sets-up an environment for the virtual machines to utilize the FPGA accelerators.

Component-wise, the Virtio-FPGA solution consists of the following elements:
\begin{itemize}
  \item A QEMU virtual device for FPGA programming
  \item A guest kernel driver to interface with the QEMU device (Virtio FPGA Manager)
  \item An extension to QEMU VFIO for ARM during Device Tree setup
  \item Proper Device Tree Overlay blob files
\end{itemize}

In our prototype, the solution targets the Xilinx Zynq UltraScale+ MPSoC for the testing and evaluation purposes. The Virtio-FPGA components are going to be detailed on the following subsections.

\subsection{FPGAs configuration}

In its internals, Virtio-FPGA implements the Virtio specification, to expose a QEMU virtual device for FPGA programming to the guests. The virtual device has the role of fulfilling FPGA programming requests that arrive from guest systems. By interfacing with the virtual device, guests are able to trigger FPGA programming requests towards the FPGA hardware. 

\begin{figure}[htbp]
\centerline{\includegraphics[scale=0.4]{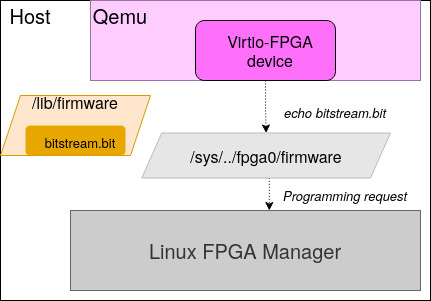}}
\caption{The interface between QEMU Virtio-FPGA device and the Linux FPGA Manager core.}
\label{fig:fig2}
\end{figure}

Fundamentally, the virtual device realizes the programming operation by making use of the Linux FPGA Manager API of the host kernel. The interface between the virtual device and the Linux FPGA Manager core, resides on the sysfs firmware entry that is available on the Xilinx Linux FPGA Manager implementation. Through this sysfs entry, it is possible for user-space processes to trigger FPGA programming requests by providing the bitstream configuration file name. Also, the requirement is that the bitstream file should be present under the \texttt{/lib/firmware} directory, in the host system. The interface between the QEMU Virtio-FPGA device and the Linux FPGA Manager is better depicted on figure \ref{fig:fig2}.

So, given an amount of bitstream files present under the host's \texttt{/lib/firmware} directory, guests can choose any of the available configurations to program FPGAs with. In a possible future Virtio-FPGA extension, the binary data of the image might be also communicated between the guest and the host system, using scatter-gather lists. 

For what concerns the guest interface, a corresponding Virtio-FPGA driver is loaded in the guest kernel, responsible to manage the paravirtualized QEMU device. The main design objectives of the guest kernel driver are:

\begin{itemize}
  \item To provide transparency, e.g. guests to function as if they were a native system
  \item To expose an interface for user-space guest requests
  \item To communicate with the host introducing a minimum overhead
\end{itemize}

The Virtio-FPGA solution aims to provide transparency in the guest system. In fact, the guest FPGA programming interface is built on top of a new, Virtio FPGA Manager that implements the FPGA Manager set of operations. Kernel solutions in the guest can take advantage of the Virtio FPGA Manager and interact with it, to program the FPGAs given a hardware configuration image. 

Similarly to the host-side FPGA programming interface, the guest kernel driver introduces a sysfs entry for triggering the programming requests from a user-space context. So, as soon as the driver is inserted into the guest system, the procedure to carry-out an FPGA programming request is as follows:

\texttt{echo filename >\\
    /sys/class/virtio\_fpga/fpga0/firmware}

For what concerns the filename, as already mentioned, it should be present under the host \texttt{/lib/firmware} directory. Then, once the programming operation terminates, a kernel message is generated to report the status of the FPGA device (whether the operation was successful or not).

At its base, the Virtio FPGA Manager communicates with the host the necessary information for configuring the FPGAs. With the Virtio protocol, the guest-host communication takes place through shared memory regions. In our scenario, the shared memory regions are two virtqueues; one containing the image filename residing on the host system, and one ready to accept the status of the programming operation, once it is over. More in details and with focus on the introduced overhead for FPGAs configuration, a context-switch to kernel space and a memory copy are needed to prepare the data for host consumption. Thus, the configuration overhead is insignificant and for this reason, it is not being considered during the evaluation procedure. 

\begin{figure}[htbp]
    \centerline{\includegraphics[scale=0.3]{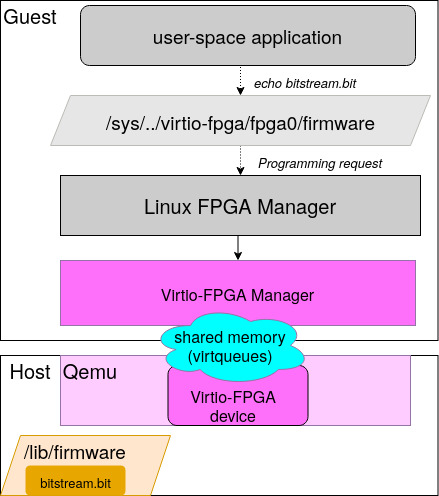}}
    \caption{The internals of the guest interface with the QEMU Virtio-FPGA device.}
    \label{fig:fig3}
\end{figure}

Figure \ref{fig:fig3} provides an overview of the components involved, for a guest system to interact with the QEMU Virtio-FPGA device, in order to trigger FPGA programming requests.

\subsection{FPGA resources utilization}

As soon as the FPGA is configured, the guest is able to utilize the resources by making use of QEMU VFIO-passthrough and Device Tree Overlays technologies.

For the Virtio-FPGA solution to function, an extension on QEMU device tree setup was required to open-up the range of the supported VFIO platform devices. 
Specifically, a generic device tree node configuration was introduced providing basic properties (e.g. interrupts, memory areas) to expose the device in the guest system. Additionally, the VFIO platform driver in QEMU does not support hotplugging of devices, so all candidate FPGA devices introduced through VFIO during QEMU bootup are initially disabled, and remain at this state until an FPGA configuration request takes place. 

For our virtualization purposes, Device Tree Overlays are the key to provide a dynamic solution to guest systems with no hotplugging support. This technology is used in Virtio-FPGA in order to dynamically expose FPGA devices to guest systems after their hardware configuration is carried-out. 

More in details, the dynamic exposure of devices to guests happens as follows: First, the guest Linux kernel is compiled with Device Tree Overlays support and user-space support for overlays application. Then, QEMU boots, passing-through some candidate FPGA devices with most basic properties to the guest system. These FPGA devices are not yet part of the FPGA hardware, but they might become upon some future configuration. The requirement for this step, is that the configurations applied should be respected throughout the lifetime of the host system. To explain more, the basic properties (e.g. physical memory areas) for the FPGA devices that are not yet physically part of the hardware, are described early in the host software and become bound to the VFIO and IOMMU subsystems. This information must be strictly followed by all the firmware images that might be used for FPGAs configuration, upon a guest request.

As a next step, the guest can trigger FPGA programming requests to the Virtio-FPGA and then apply a device tree overlay in the guest system to enable the corresponding devices. During this procedure, any other device-specific information can be provided in the overlay (e.g. clocks, compatible strings etc.).

The use of device tree overlays is necessary in systems where information about non-present hardware needs to be provided to guests in advance. In the future, this workaround might be obsolete, in case that hot plug support is added to platform devices in QEMU.

\section{Benchmarks}

In this section, we present the evaluation procedure and the experimental results of the Virtio-FPGA prototype. The main objective of the evaluation is to demonstrate the virtualization overhead of the presented solution. As already mentioned, the virtualization overhead for FPGAs configuration is minimal, and thus it is omitted from the evaluations. Thus, we will measure the virtualization overhead introduced during FPGAs utilization in the scope of the Virtio-FPGA solution.

To do this, we measured and compared the performance of interfacing with the FPGA hardware from a guest system and the host system.
All the measurements were performed on a Xilinx Zynq Ultrascale+ MPSOC (ZCU102), which features a quad-core ARM Cortex-A53, with 4GB DDR4 memory attached to the PS (Processing System) and 512MB DDR4 memory attached to the PL (Programmable Logic). 
The host system is built using the Petalinux tools v2021.2 with Xilinx Linux version 5.10.0, whereas the guest system is a buildroot-based arm64 QEMU virt model with Linux version 6.0. 

 To measure the virtualization overhead of FPGAs utilization we setup a bitstream file that instructs the FPGAs to feature a CDMA (Central Direct Memory Access) device, capable of carrying-out DMA transfers between areas of the on-chip DDR memory. Then, we applied this configuration in the Virtio-FPGA solution, in a way that the CDMA device of the PL is made available to the guest system. 

Given this setup, we ran a DMA test benchmark from the guest, and obtained performance results for the DMA transfers. Last, we compared the results with a scenario where the DMA test benchmark is executed directly in the host system.

\subsection{Hardware benchmark application}
\begin{figure}[htbp]
\centerline{\includegraphics[scale=0.14]{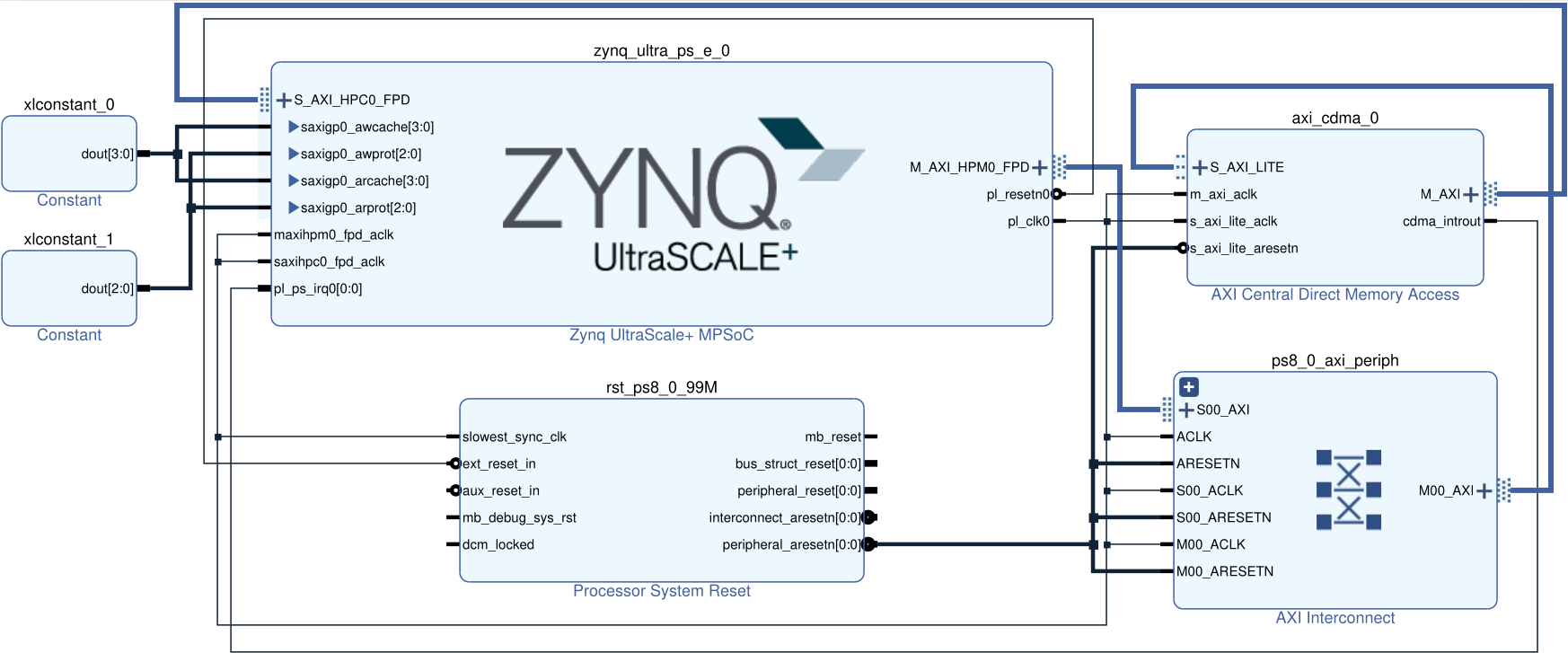}}
\caption{The design for the evaluation part as generated by Vivado 2021.2.}
\label{fig:fig5}
\end{figure}

To further detail the hardware configuration of the FPGAs, we generated the bitstream file using Vivado 2021.2. In order to prepare the configured CDMA IP for being passed through with QEMU VFIO, we had to ensure from the design, I/O Coherency between the PL and the PS. For this reason, we used the HPC0 port of the AXI, to connect the CDMA IP to the PS. This is because with the HPC0 port, it is possible for coherent masters in the PL to snoop the CPU caches, given also that the AxCACHE signals of the port are properly configured. As an additional requirement, the CCI (Cache Coherent Interconnect) should be configured to enable snooping, and we addressed this given that we used the ATF (Arm Trusted Firmware) binary on the host, which configures the CCI by default.  Also, we had to program the \texttt{LPD\_APU} register of the \texttt{LPD\_SLCR} module, to enable the broadcasting of CPU caches to the PL masters. We did this, by setting the register from the boot-up script of the host system. Last but not least, we setup the AxPROT signals of the HPC0 port, so that accesses to the CDMA IP from EL1 (e.g. Linux accesses) are allowed. In figure \ref{fig:fig5} we present the final design used for the hardware configuration of the FPGAs in this evaluation.

\subsection{Software benchmark application}
\begin{figure}
\centerline{\includegraphics[scale=0.45]{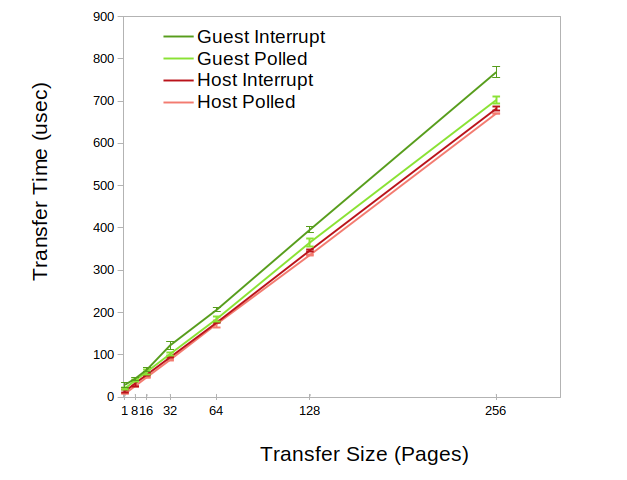}}
\caption{The average transfer time and the standard deviation in respect to the transfer size, for host and guest setups with interrupt and polled methods.}
\label{plot1}
\end{figure}

The DMA test benchmark is based on the DMA test client, which is available in the upstream Linux kernels and can be enabled with CONFIG\_DMATEST option during kernel compilation.
For our evaluation purposes, a few tweaks were applied on the default DMA test client, so that we obtain the performance numbers. 
We performed DMA transfers of various memory sizes, ranging from 1 Page (meaning, 4KB data) to 256 Pages transfer size. Also, we carried-out both interrupt-based and polled-based transfers. As for the evaluation numbers, we measured:

\begin{enumerate}
    \item \textbf{The preparation time:} Time (in usec) needed to map the memory ranges to the DMA address space, to be prepared for the DMA transfer.
    \item \textbf{The transfer time:} Time (in usec) needed from when we program the CDMA device to start the transfer, till it ends. We evaluated both:
    \begin{enumerate}
        \item {The interrupt transfer time}
        \item {The polled transfer time}
    \end{enumerate}
\end{enumerate}
For each transfer size, we ran 500 iterations, and extracted the average and standard deviation values of the preparation and transfer times. For the preparation times, the results that we obtained are presented in table \ref{tab1}. Also, figure \ref{plot1} graphically demonstrates the average value for the transfer time, along with the standard deviation value, in respect to the transfer size, for the two systems and the two transfer methods.

\begin{table}[htbp]
\caption{Preparation time (usec)}
\begin{center}
\begin{tabular}{|c|c|c|c|c|c|c|}
\hline
&\multicolumn{6}{|c|}{\textbf{Transfer size (4KB Pages)}} \\
\cline{2-7}
& \textbf{\textit{1}}& \textbf{\textit{16}}& \textbf{\textit{32}}&\textbf{\textit{64}}& \textbf{\textit{128}}& \textbf{\textit{256}} \\
\hlinewd{1pt}
\textbf{Host avg*}& 6.14 & 62.5 &  122.3 & 242.37 & 474.334 & 748.6\\
\hline
\textbf{Host std}& 1.79 &  1.53 &  1.54 & 2.42 & 6.11 & 7.74\\
\hlinewd{1pt}
\textbf{Guest avg}& 4.36 & 42.34 & 82.79 & 162.8 & 292.18 & 411.2\\
\hline
\textbf{Guest std}& 7.65 & 8.84 & 7.98 & 8.56 & 14.59 & 19.66\\
\hline
\multicolumn{4}{l}{$^{\mathrm{a}}$All results are in usec}
\end{tabular}
\label{tab1}
\end{center}
\end{table}

\subsection{Results}
From the results, at first we see that the preparation time is always worse in the native versus the virtualization setup. This behavior can be explained when looking into how IOMMU translations are generated. In the virtualization case, the guest system uses a direct model for the DMA mappings, so that it seemingly programs the DMA using guest physical addresses. All the necessary IOMMU translations have been already setup for the CDMA IP, thanks to the VFIO platform driver of QEMU. From the other hand, in the native case the IOMMU translations are created upon DMA mapping events, and thus no direct model is being utilized. So, for each DMA mapping event, the native case has to pay the additional overhead for seting up the translations for the IOMMU subsystem.

With respect to the results obtained for the transfer times, for both guest and host setups we observe that for the same transfer size, the polled transfer time is always better than the interrupt transfer time. This is normal, considering that the transfer sizes are relatively small, and thus it is not worthy to put the execution thread to sleep and wait for the interrupt to occur. Related to the standard deviation values, the guest metrics usually tend to be more unpredictable than the host, as the guest environment has to deal with its threads being scheduled-out from time to time. In any case, the highest standard deviation value is noticed for a guest transfer of 1MB data, with a value of 13.28 usecs.

Considering the virtualization overhead occurring during the transfer, it is greatly influenced by the transfer size of the DMA operation.
In the polled case, the overhead is decreasing from a value of 152\%, for 4KB data, down to a value of 5\%, for 1MB data. In the interrupt case, the higher value for the overhead is 112\% for 4KB data, whereas the lower value is 12.6\% for 1MB data.
In all, we conclude that the larger the amount of data to be transferred, the more the virtualization overhead is hidden. For reasonable transfer sizes (e.g. 256KB - 1MB) the average overhead for the interrupt case is 15\%, whereas for the polled case is 7\%, which is normally lower given the cost of interrupts virtualization.

\section{Related Work}
In the research community, there exists a wide range of academic studies regarding FPGA resources virtualization. The existing literature in the field varies, from PCIe and network attached, to SoC-attached FPGAs.

For PCI-e and network-attached FPGAs, there is a plethora of works that aim to provide FPGAs virtualization. Epigrammatically,  the FPGAVirt \cite{b9} solution exploits the Virtio-Vsock layer to enable communication between guest systems and network-attached FPGAs. Another interesting solution is pvFPGA \cite{b10}, which is based on the Xen hypervisor to provide a paravirtualized environment for PCI-e attached FPGAs. Also, with focus on the scalable management of homogeneity of the FPGA hardware in the cloud, RC3E \cite{b11} presents a hypervisor that provides an interface for guest systems to interact with virtual FPGA components.

In the embedded domain, there are few works that provide virtualization of the
FPGA resources with microkernel hypervisors. The work in \cite{b12} ports the CODEZERO hypervisor to a platform with SoC-attached FPGA, involving necessary extensions to support FPGA hardware tasks management and scheduling. Also, of much interest is the work in \cite{b13}, which ports the NOVA hypervisor to an embedded platform with attached FPGA, in a way that it can delegate the management of DPR (Dynamic Programmable Regions) resources. In the same concept, the work in \cite{b14} builds a custom hypervisor called Ker-one, which leverages the FPGA DPR regions as virtual devices to guest systems. Then, the hypervisor is able to identify and dynamically handle the programming requests that target the DPR regions.
These works deal indeed with many challenges that FPGAs virtualization present. However, the solutions presented are much platform-specific, since there is no unified API to carry-out the programming of FPGA resources for multiple boards. In our solution, the virtualization of the Linux FPGA Manager creates that unified API which hides the manufacturer details from the upper software layers.

\section{Conclusion}

In this work we presented Virtio-FPGA: a new virtualization solution for SoC-attached FPGAs.
Its main contribution is that it proves it is feasible to dynamically re-configure and access FPGA resources, based entirely on Linux interfaces and APIs. It is a solution that is flexible, in virtue of interfacing with the generic Linux FPGA Manager API to virtualize the configuration of FPGAs, and also efficient, as a consequence of utilizing Virtio and VFIO technologies to eliminate the virtualization overhead. The virtualization overhead is evaluated with an example CDMA FPGA device, being no higher than 12\% for at least 1MB DMA transfers. Future works include adding support for partially reconfiguring the FPGA regions, as well as integrating a monitor tool in the hypervisor to keep track of the hardware tasks.




\begin{thebibliography}{00}
\bibitem{b1} Bouhali, Mustapha, Farid Shamani, Zine Elabadine Dahmane, Abdelkader Belaidi, and Jari Nurmi. "FPGA applications in unmanned aerial vehicles-a review." In International Symposium on Applied Reconfigurable Computing, pp. 217-228. Springer, Cham, 2017.
\bibitem{b2} F. Eberli, "Next Generation FPGAs and SOCs - How Embedded Systems Can Profit," 2013 IEEE Conference on Computer Vision and Pattern Recognition Workshops, 2013, pp. 610-613, doi: 10.1109/CVPRW.2013.92.
\bibitem{b3} M. H. Quraishi, E. B. Tavakoli and F. Ren, "A Survey of System Architectures and Techniques for FPGA Virtualization," in IEEE Transactions on Parallel and Distributed Systems, vol. 32, no. 9, pp. 2216-2230, 1 Sept. 2021, doi: 10.1109/TPDS.2021.3063670.
\bibitem{b4} R. Russell, “Virtio: towards a de-facto standard for virtual i/o
devices,” ACM SIGOPS Operating Systems Review, vol. 42, no. 5, pp. 95–103, 2008
\bibitem{b5} Alan Tull, "Reprogrammable Hardware under Linux", Presentation at Embedded Linux Conference Europe 2015, Oct. 2015, [online] Available: http://events.linuxfoundation.org/sites/events/files/slides/FPGAs-under-Linux-Alan-Tull-vl.00.pdf
\bibitem{b6} Yassour, Ben-Ami \& Ben-Yehuda, Muli \& Wasserman, Orit. (2008). "Direct device assignment for untrusted fully-virtualized virtual machines." IBM Research Division Haiafa Research Laboratory, September 20, 2008 
\bibitem{b7} Williamson, Alex, Alexey Kardashevskiy, Linus Torvalds, Zi Shen Lim, Gavin Shan, and Mauro Carvalho Chehab. "VFIO-” Virtual Function I/O”." O” https://www. kernel. org/doc/Documentation/vfio. txt (2017).
\bibitem{b8} A. Motakis, A. Rigo and D. Raho, "Platform Device Assignment to KVM-on-ARM Virtual Machines via VFIO," 2014 12th IEEE International Conference on Embedded and Ubiquitous Computing, 2014, pp. 170-177, doi: 10.1109/EUC.2014.32.
\bibitem{b9} J. Mbongue, F. Hategekimana, D. Tchuinkou Kwadjo, D. Andrews and C. Bobda, "FPGAVirt: A Novel Virtualization Framework for FPGAs in the Cloud," 2018 IEEE 11th International Conference on Cloud Computing (CLOUD), 2018, pp. 862-865, doi: 10.1109/CLOUD.2018.00122.
\bibitem{b10} Wei Wang, M. Bolic and J. Parri, "pvFPGA: Accessing an FPGA-based hardware accelerator in a paravirtualized environment," 2013 International Conference on Hardware/Software Codesign and System Synthesis (CODES+ISSS), 2013, pp. 1-9, doi: 10.1109/CODES-ISSS.2013.6658997.
\bibitem{b11} O. Knodel, P. Lehmann and R. G. Spallek, "RC3E: Reconfigurable Accelerators in Data Centres and Their Provision by Adapted Service Models," 2016 IEEE 9th International Conference on Cloud Computing (CLOUD), 2016, pp. 19-26, doi: 10.1109/CLOUD.2016.0013.
\bibitem{b12} K. Dang Pham, A. K. Jain, J. Cui, S. A. Fahmy and D. L. Maskell, "Microkernel hypervisor for a hybrid ARM-FPGA platform," 2013 IEEE 24th International Conference on Application-Specific Systems, Architectures and Processors, 2013, pp. 219-226, doi: 10.1109/ASAP.2013.6567578.
\bibitem{b13} T. Xia, J. -C. Prevotet and F. Nouvel, "Mini-NOVA: A Lightweight ARM-based Virtualization Microkernel Supporting Dynamic Partial Reconfiguration," 2015 IEEE International Parallel and Distributed Processing Symposium Workshop, 2015, pp. 71-80, doi: 10.1109/IPDPSW.2015.72.
\bibitem{b14} Tian Xia, M. -A. -F. Rihani, J. -C. Prévotet and F. Nouvel, "Demo: Ker-ONE: Embedded virtualization approach with dynamic reconfigurable accelerators management," 2016 Conference on Design and Architectures for Signal and Image Processing (DASIP), 2016, pp. 225-226, doi: 10.1109/DASIP.2016.7853825.
\bibitem{b15} C. Pilato et al., "EVEREST: A design environment for extreme-scale big data analytics on heterogeneous platforms," 2021 Design, Automation \& Test in Europe Conference \& Exhibition (DATE), Grenoble, France, 2021, pp. 1320-1325, doi: 10.23919/DATE51398.2021.9473940.
\end{thebibliography}
\end{document}